\documentclass[pra,
 reprint,
superscriptaddress,
 amsmath,amssymb,
 aps,
]{revtex4-2}

\usepackage{graphicx}
\usepackage{dcolumn}
\usepackage{bm}
\usepackage{xcolor}
\usepackage{comment}

\begin{document}

	\title{Discriminating the phase of a coherent tone with a flux-switchable superconducting circuit}
	
	\author{L. Di Palma}
	\affiliation{Dipartimento di Fisica ``Ettore Pancini", Università degli Studi di Napoli Federico II, I-80125 Napoli, Italy}
 \affiliation{SEEQC, Corso Protopisani 70, Napoli 80146, Italy}
	
	\author{A. Miano}
	\affiliation{Department of Applied Physics, Yale University, New Haven, Connecticut 06520, USA}
	
	\author{P. Mastrovito}
	\affiliation{Dipartimento di Fisica ``Ettore Pancini", Università degli Studi di Napoli Federico II, I-80125 Napoli, Italy}
	
	\author{D. Massarotti}
	\affiliation{Dipartimento di Ingegneria Elettrica e delle Tecnologie dell’Informazione,
Università degli Studi di Napoli Federico II, via Claudio, I-80125 Napoli, Italy}
\affiliation{CNR-SPIN, UOS Napoli, Monte S. Angelo, via Cinthia, I-80126 Napoli, Italy}

	\author{M. Arzeo}
	\affiliation{SEEQC, Corso Protopisani 70, Napoli 80146, Italy}
	
	\author{G. P. Pepe}
	\affiliation{Dipartimento di Fisica ``Ettore Pancini", Università degli Studi di Napoli Federico II, I-80125 Napoli, Italy}

	\author{F. Tafuri}
	\affiliation{Dipartimento di Fisica ``Ettore Pancini", Università degli Studi di Napoli Federico II, I-80125 Napoli, Italy}
	\affiliation{CNR - Istituto Nazionale di Ottica (CNR-INO), Largo Enrico Fermi 6, 50125 Florence, Italy}
	\author{O. Mukhanov}
	\affiliation{ SEEQC, Elmsford, NY 10523, USA}

\begin{abstract}
We propose a new phase detection technique based on a flux-switchable superconducting circuit, the Josephson Digital Phase Detector (JDPD), which is capable of discriminating between two phase values of a coherent input tone. 
When properly excited by an external flux, the JDPD is able to switch from a single-minimum to a double-minima potential and, consequently, relax in one of the two stable configurations depending on the phase sign of the input tone. \newline
The result of this operation is digitally encoded  in the occupation probability of a phase particle in either of the two JDPD wells.
In this work, we demonstrate the working principle of the JDPD up to a frequency of 400 MHz  with a remarkable agreement with theoretical expectations. As a future scenario, we discuss the implementation of this technique to superconducting qubit readout. We also examine the JDPD compatibility with the Single Flux Quantum (SFQ) architecture, employed to fast drive and measure the device state.
\end{abstract}

\maketitle

\section{Introduction}

The detection of coherent tones is a crucial operation in many experiments, ranging from the detection of elusive particles \cite{axion1,axion2,axion3,Tsuji} to quantum information processing \cite{GU20171_photonics,Siddiqi_heterodyme,Mallet_heterodyme,Martins_heterodyme}. In the last decades, the development of circuit Quantum Electrodynamics (QED) \cite{RevModPhys.93.025005_qed,blais_heterodyme}, based on superconducting devices, led to a growing interest in measurements of microwave signals, which typically involve the use of superconducting cavities \cite{GU20171_photonics,RevModPhys.84.1_nori_detect}. As an example of application, the readout of a superconducting qubit is commonly performed with a Quantum Non-Demolition (QND) operation which relies on the frequency response of a co-located resonator, being dependent on the state of the qubit via dispersive coupling \cite{blais_heterodyme}. The cavity can be probed in reflection with a tone at a fixed frequency, encoding the state of the qubit in the relative phase and amplitude of the reflected pulse \cite{Siddiqi_heterodyme, Mallet_heterodyme, martinis_surface_code2, Martins_heterodyme}. The first experiments that successfully performed the dispersive readout of a qubit with acceptable fidelities were designed to encode the information in the relative amplitude of the reflected pulse \cite{blais_heterodyme}. Later, the advent of Josephson parametric amplifiers \cite{single_shot_Krantz} made it possible to sense the phase response of the readout resonator with fidelities above 99\%. \newline
In the last several years, there was a growing interest in superconducting digital circuits \cite{Opremcak2,Opremcak,mcdermot_sfq}, which operate with low power dissipation at cryogenic temperatures\cite{x1,x2,x3}. These devices have been proposed as control and readout interfaces for cryogenic circuits \cite{Takeuchi_2019,qubit_control_mcdermott} with the goal to reduce hardware complexity and improve system scalability \cite{oleg1}. The state of the art digital readout of microwave photons relies on the Josephson Photo Multiplier (JPM), which operates at the mK stage of a dilution refrigerator \cite{Govia,Opremcak,Opremcak2}. The JPM consists of a bistable superconducting circuit whose operating point can be made sensitive to the average number of photons in the readout cavity. This information is mapped in the current direction flowing through the device \cite{Opremcak,Opremcak2}. A possible way to read the JPM's state exploits propagating fluxons \cite{caleb1} produced by a Single Flux Quantum (SFQ) classical circuit \cite{mcdermot_sfq}. In this framework, JPM is proposed as part of a more complex architecture, in which classical qubit control, measurement, and data processing are performed by an SFQ processor. This approach can be a valid solution to enhance the scalability of superconducting quantum systems, which remains a big engineering challenge to realize practical error-corrected quantum computers \cite{oleg1}. \newline
Despite the advantages of such a detection technique, the JPM approach still requires an adiabatic externally generated flux bias in order to tune the device during the many steps necessary for successful  Quantum Non-Demolition (QND) \cite{QND} readout. Measurement is performed in resonance with the  cavity and requires non-trivial operations to reduce the backaction on the quantum chip,  which limits the readout speed. Moreover, the JPM protocol provides an intrinsically state-dependent fidelity that can be harmful in some experiments \cite{Opremcak,Opremcak2}.

In this article, we propose a new phase detection technique based on a  flux-switchable superconducting circuit, which is capable of discriminating between two phase values of a coherent input tone. For instance, such a tone can be a readout pulse whose phase encodes the outcome of a qubit measurement. The device operation is compatible with SFQ circuits, which could be employed to perform the detection in a fully digital fashion. Our device is based on a bistable Quantum Flux Parametron (QFP) \cite{hosoya1991quantum}\cite{x6}, which has already been proposed for classical \cite{Takeuchi2014Sep,Takeuchi2017Mar,Tsuji} and quantum applications, including superconducting flux qubits controlled with fast pulses \cite{Poletto_2009,Castellano_2010,Chiarello_2012} or  ultralow-power logic device \cite{x6}. 
In this technique, the QFP is at first flux-biased into a harmonic configuration to sense the input coherent tone, then quickly flux-switched to a bistable configuration to store the information on the tone's phase. We name our device \emph{Josephson Digital Phase Detector} (JDPD), since the result of the detection is naturally encoded in the occupation probability of a phase particle in either of two wells in the bistable configuration. 

In this paper, we present a circuit model of the JDPD and discuss the operating principle, with some consideration on the symmetries required to achieve high performances and suppress the backaction due to the diabatic flux-switch. Numerical simulations with QuTip \cite{QuTip} are performed, showing how the JDPD can perform phase detection with a fidelity of 99.99$\%$ when flux-switched on a timescale of 0.05 ns. From these results, we discuss the compatibility of the JDPD with SFQ technology.
We also present experimental results on a preliminary device, which is capable of detecting the phase of a  400 MHz input coherent tone with a 99.98 $\%$ fidelity when flux-switched on a timescale of 1 ns. The acquired data confirm the feasibility of the operating principle and are in very good agreement with numerical simulations adapted to our experimental scenario.

For future prospects, we discuss how we are currently implementing the next generation of experimental devices, which would be flux-switched by an integrated pulse generator based on SFQ logic to perform fast phase detection of a GHz input tone. SFQ circuits can also be employed to generate an input signal phase locked to the flux-switching pulse. We also address the compatibility with the JDPD approach applied to superconducting qubits readout and we envision its employment as digital readout modules in SFQ-based quantum information processing platforms, where the JDPD output is converted in SFQ form.

\section{Device model}
\begin{figure}
    \centering
\includegraphics{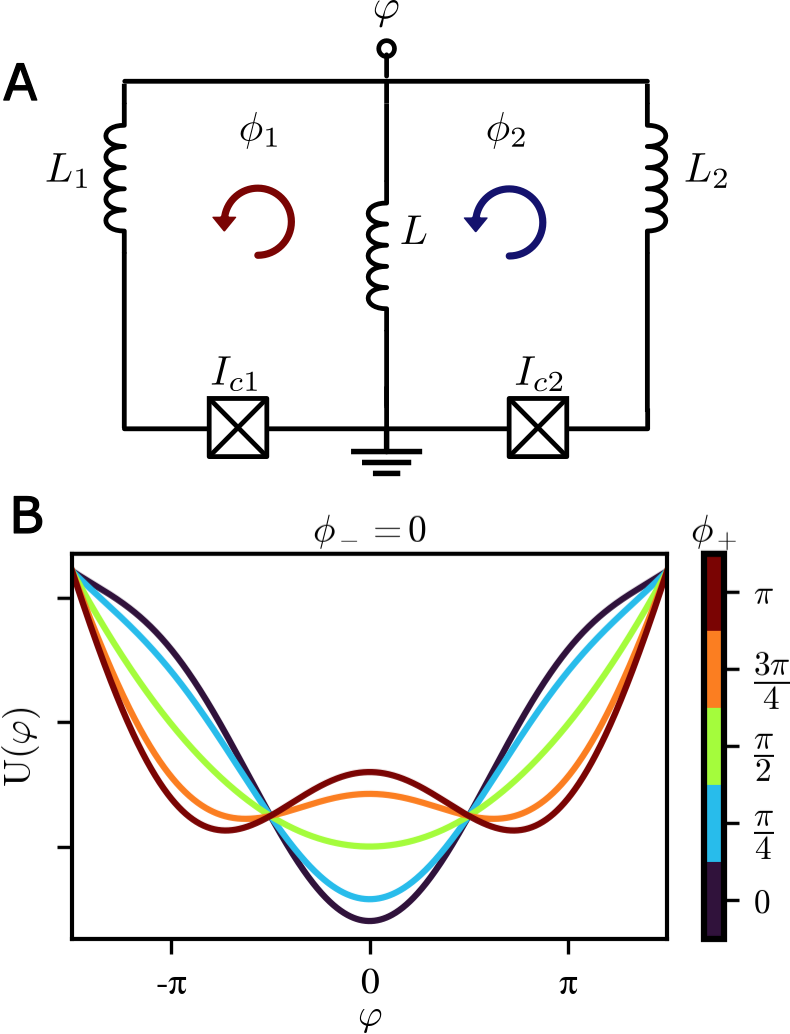}
    \caption{Schematic of the Josephson Digital Phase Detector (JDPD) (A).
    The JDPD is composed of a nominally symmetric Quantum Flux Parametron (QFP), whose left and right loops are phase-biased by independent phases $\phi_{1(2)}$. The bias phases are related to the magnetic fluxes $\Phi_{1(2)}$ threaded to each loop by the relation $\phi_{1(2)} = 2\pi\frac{\Phi_{1(2)}}{\Phi_0}$.
    (B). Potential energy landscapes of the JDPD for different values of common flux $\phi_+ = (\phi_1 + \phi_2)/2$, for $\phi_- = (\phi_1 - \phi_2)/2 = 0$. The potential energy can present one absolute minimum (blue curve), harmonic potential landscape (green curve) or two absolute minima (red curve), depending on the external magnetic flux.}
    \label{fig:fig_1}
\end{figure}

The fundamental block of a JDPD is based on a QFP (Fig. 1A), whose left and right loops can be independently flux-biased by, respectively, the normalized fluxes $\phi_1$ and $\phi_2$:
\begin{equation}
\phi_{1,2} = 2 \pi\frac{\Phi_{1,2}}{\Phi_0} 
\end{equation}
where $\Phi_{1,2}$ is the magnetic flux threaded, respectively, to the left and the right loops of the circuit in Fig. \ref{fig:fig_1} A.
To describe the flux-tuning properties of the device in a compact form, it is useful to introduce the common and differential fluxes $\phi_+$ and $\phi_-$, defined as
\begin{equation}
\begin{gathered}
 \phi_{+} = \frac{\phi_1 + \phi_2}{2} \\
 \phi_{-} = \frac{\phi_1 - \phi_2}{2}.
\end{gathered}
\end{equation}
These can be independently generated by properly designed flux lines, as shown later in the experimental section of the manuscript. In the symmetrical case where I$_{c1}$ = I$_{c2}$ = I$_c$, and neglecting the energy stored in the stray linear inductances under the assumption L$_{1,2}<$L$_{J}$, the potential energy function of the device is given by:
\begin{equation}
\label{eqn:potential_symm}
    \frac{U(\varphi)}{E_\mathrm{L}}  = \frac{\varphi^2}{2} - 2 \beta_L \cos( \phi_+) \cos(\varphi+ \phi_-)
\end{equation}
where $\beta_L=2 \pi I_cL/\Phi_0$ and $\varphi$ is the phase drop across the central inductor $L$.
The potential energy in Eq. \eqref{eqn:potential_symm} is normalized to the linear inductor energy $E_\mathrm{L} = \left(\frac{\Phi_0}{2\pi}\right)^2/L$. This expression is formally identical to the potential energy of a double-SQUID \cite{Poletto_2009,Castellano_2010,Chiarello_2012}.
If $2\beta_L \leq 1$ the potential has only one absolute minimum for any combination of $\phi_{\pm}$. If $2\beta_L>1$, the bias fluxes can be tuned to generate potential shapes with multiple minima. For instance, in Fig. 1(b), for a device with $2\beta_L = 3$ and for $\phi_- = 0$, the potential energy can host a single absolute minimum for $\phi_+ = 0$ or two degenerate minima for $\phi_+ = \pi$. The device can also be flux-biased to a harmonic configuration for $\phi_+ = \pi/2$, where the nonlinear term in Eq. \eqref{eqn:potential_symm} is suppressed. The detection technique exploits the  tunability of the JDPD's potential energy to perform phase detection.

The kinetic energy is given by capacitive contributions of the Josephson junctions inside the rf loops, i.e:

\begin{equation}
\label{kinetic_energy}
    T = 2 \frac{e^2}{2C} n^2 = 2E_c n^2  
\end{equation}

where C$_1$=C$_2$=C is assumed for both junction capacitances and $n$ is the operator number.
Assembling equations \ref{eqn:potential_symm} and \ref{kinetic_energy}, the static JDPD Hamiltonian reads as:
\begin{equation}
    \label{hamiltonian}
    H = 2 E_c n^2 + E_L\left[\frac{\varphi^2}{2} - 2 \beta \cos\phi_+ \cos\left(\varphi+ \phi_-\right)\right].  
\end{equation}

\section{Principle of operation}
\begin{figure*}
\includegraphics{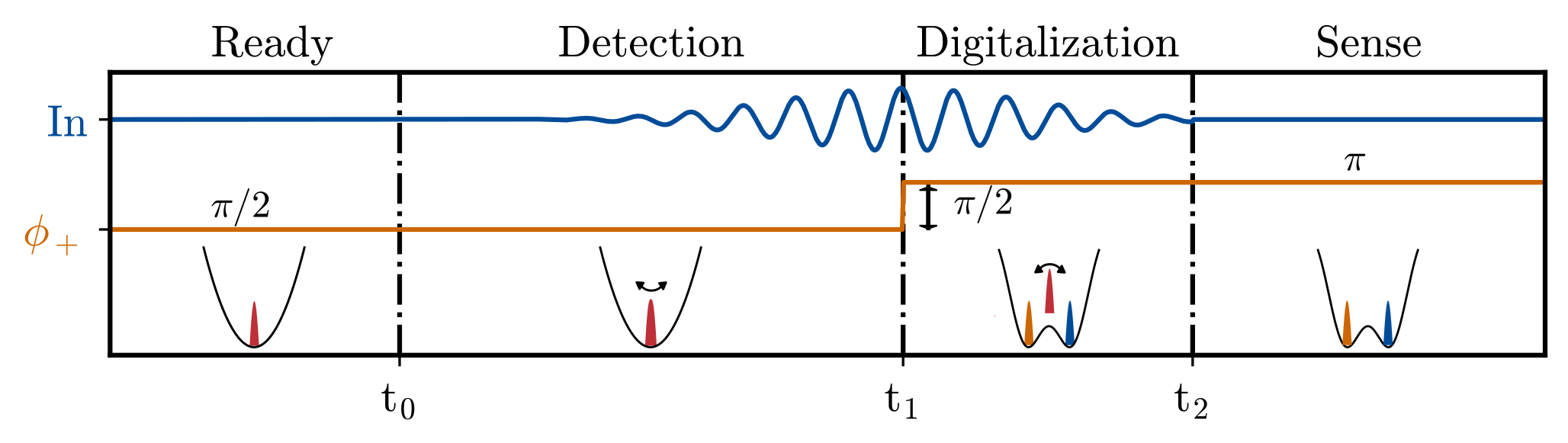}

\caption{Timing diagram of the phase detection performed with the JDPD. The detector is initially reset in the ground state of the harmonic configuration by applying $\phi_+ = \pi/2$.  In the ``Detection" step, the JDPD gets in touch with the input tone to be detected. The device's wavefunction oscillates around the potential minimum following the stimulus. The ``Digitalization" is accomplished by flux-switch the JDPD in the double well configuration. The wavefunction will collapse in the left $|L>$ or right state $|R>$ depending on the phase displacement of the input tone. In the ``Sense" step, we probe the JDPD to gather information on the wavefunction's final state.} 
\label{fig:2panel}
\end{figure*}
The diagram of the phase detection protocol is shown in Fig. \ref{fig:2panel}. The protocol is composed of four separate steps, respectively labeled as ``Ready", ``Detection", ``Digitalization" and ``Sense". The JDPD is first configured in a harmonic shape, by providing $\phi_+ = \pi / 2$, as shown in the first part of Fig. \ref{fig:2panel}. The Hamiltonian in Eq. \eqref{hamiltonian} takes the form:
\begin{equation} 
    H_\mathrm{ho} = 2 E_C n^2 + E_L \varphi^2  = \hbar \omega_0 \left(\frac{1}{2} + a ^ \dagger a\right)
\label{hamiltonian_ho}
\end{equation}
where $\omega_0= \sqrt{ 4 \;E_C \; E_L } /\hbar $ is the natural frequency of the equivalent harmonic oscillator. In the absence of any external perturbation, we expect the wavefunction to relax in the ground state of the potential, which makes the device ``Ready" to start phase detection. \newline
In the ``Detection" step of Fig. \ref{fig:2panel}, an input current stimulus  $I(t)$ is applied through the JDPD input node. The system evolves according to the  Hamiltonian of a driven harmonic oscillator,
\begin{equation}
    H_\mathrm{tot} =  H_\mathrm{ho} - \frac{\Phi_0 I(t)}{2 \pi} \varphi.
\end{equation}
For simplicity, we consider the case in which  $I(t)$ is a sinusoidal tone with amplitude $I_0$, angular frequency $\omega_r$ and displacement  $\theta_r$ :
\begin{equation}
    I(t) = I_0 \sin\left(\omega_r t + \theta_r\right).
\end{equation}
The wavefunction is described by a coherent state, in which the expectation value of ${\varphi}$ follows the input tone's time evolution:
\begin{equation}
    <{\varphi}> (t) = \varphi_0\sin\left(\omega_{r} t + \theta_0\right)
\end{equation}
where, in general, $\varphi_0$ and $\theta_0$ depend on the $I(t)$ parameters.
For instance, in the case in which $\omega_r< \omega_0$, $\varphi_0$ and $\theta_0$ take the values: 
\begin{align}
  \varphi_0 &\approx 2\pi\frac{LI_0}{\Phi_0}\\
  \theta_0 &\approx \theta_r
\end{align}
which means that the information on the input tone displacement $\theta_r$ is transferred to $\theta_0$. \newline 
According to the quantum mechanical evolution of a coherent state, the wavefunction standard deviation \cite{driven_harm_osci} is constant in time and it is equal to:
\begin{equation}
    \sigma = 2 e \sqrt{\frac{2 Z_0}{\hbar}}
    \label{eq:stnd_dev}
\end{equation}
where $Z_0 = \sqrt{L/C}$ is the characteristic impedance of the $LC$ resonator. \newline 
The ``Digitalization" step, shown in Fig. 2 is accomplished by  diabatic flux-switch  the JDPD to a bistable potential configuration ($\phi_+ = \pi/2 \rightarrow \phi_+ = \pi$). Intuitively, according to the sign of $<{\varphi}>$ at the flux-switch time $t_1$, the wavefunction of the system will primarily be confined in either of the two wells, depending on the initial phase $\theta_r$ imposed to the coherent state by the input current $I(t)$. We define these two states $|L>$ and $|R>$, associated with the wavefunction collapsing in the left or right well of the potential energy. \newline
Right after, $I(t)$ is turned off to prevent further dynamics of the wavefunction after the ``Digitalization" is completed. A detailed explanation of how the probability of confining the wavefunction in one of the two wells depends on $\theta_r$ can be found in the section ``Numerical Analysis" of the manuscript. 
The position of the phase particle can now be ``Sensed" as the two possible outcomes of a measurement of the ${\varphi}$ operator will have opposite values, which correspond to opposite signs for the current flowing through the central inductor L, according to:
\begin{equation}
    <I> = \frac{\Phi_0}{2\pi L} <\varphi>.
\end{equation}
So far, we have explained how to map the initial phase $\theta_r$ of the input current $I(t)$ to the occupation probability of the wavefunction of the JDPD in the bistable configuration. It is thus important to describe some details of the dynamics during the flux-switch, which unveils the advantages of the JDPD protocol with respect to the one used in the JPM.

The primary difference lies in the fact that we do not require the JDPD to be in resonance with the input tone that
we want to sense. This has the tremendous advantage that the emission spectrum in the bistable configuration can be designed to be far from the absorption spectrum of any nearby circuit. 
Therefore, the emitted photons generated during the relaxation of the system during the ``Digitalization" step will likely not harm the coherence of the surrounding circuitry \cite{backaction_1,backaction_2,Opremcak}.
Moreover, the symmetric topology \cite{symmetry1} \cite{symmetric2} of the JDPD suppresses the flux-induced backaction at the active node. As any real device is going to have asymmetries because of fabrication uncertainties, we also discuss in Appendix A how to correct for such imperfection by properly adjusting the differential flux $\phi_-$ before the protocol starts. 

\section{Numerical analysis}
The timing diagram, shown in Fig. \ref{fig:2panel}  has been  validated numerically using the Lindblad master equation solver \textit{mesolve} implemented in QuTip \cite{QuTip}.
These simulations allow us to derive key features of the JDPD phase detection protocol, such as the probabilities associated with the wavefunction time evolution.\newline 
The general form of the Lindblad master equation is given by:
\begin{equation}
\label{eq:lindbland}
    \frac{d \rho}{dt} = -i\bigg[ H, \rho \bigg] + \sum_k \bigg ( L_k \; \rho \; L_k^\dagger - \frac{1}{2} \bigg \{  L_k^\dagger  L_k,  \rho \bigg \} \bigg )
\end{equation}
where $ \rho$ is the density matrix of the system, $ H$ is the Hamiltonian operator, and the $ L_k$ are the Lindblad operators \cite{open1,open2,open3,QuTip}.
To describe the time evolution of the JDPD's wavefunction, we have truncated the series in Eq. \ref{eq:lindbland} at the first order and considered as Lindbland operator $ L_1$ the quantity:
\begin{equation}
     L_1 = \sqrt{\gamma}  a
\end{equation}
where the  $\gamma$ is the dissipation rate and $a$ is the annihilation operator, defined in Eq. \ref{hamiltonian_ho}, through which the environment couples to the system. This model corresponds to a relaxation map \cite{open1,open2,open3}, which describes the JDPD's energy loss assuming the temperature of the system $T=0$ \cite{Isar2003May}. \newline
Simulations have been carried out by considering a central inductor L = 220 pH and junctions with a critical current I$_c$ = 5.4 $\mu$A. These parameters come from the experimental realization of the JDPD, which will be discussed in more detail in the Section ``Measurements" of this manuscript.
The values for $L$ and I$_c$ lead to a $2 \beta_L = 7$, following the definition in Eq. \ref{eqn:potential_symm}. This means that the Josephson contribution to the total energy is large enough to guarantee the formation of a double well potential, which is necessary to confine the wavefunction when a flux-switch is applied. To model the capacitive contribution of the Josephson Junctions (JJs), we have considered  2C = 78 fF, which is the value expected from fabrication.
\newline
\begin{figure}
    \centering
    \includegraphics{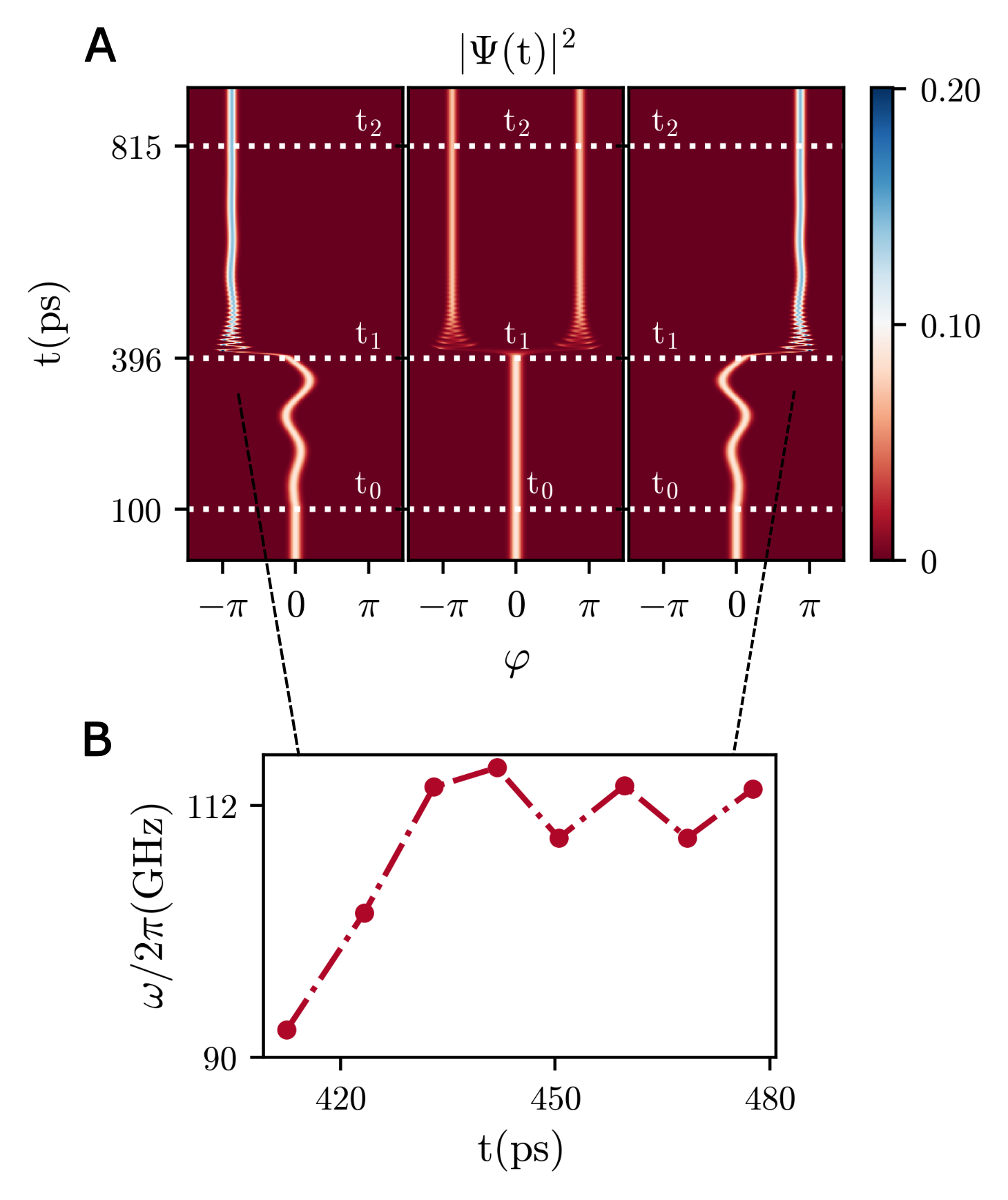}
    \caption{Simulations of JDPD with QuTip \cite{QuTip}.(A) The JDPD wavefunction has been simulated according to the timing diagram reported in \ref{fig:2panel}. In the left panel, an initial displacement $\theta_r = 0$ has been considered for the sinusoidal input tone. As consequence, the wavefunction collapses in the left well after the flux-switch tone. The same sequence has been simulated in the right panel but with an initial displacement $\theta_r = \pi$ of the input tone. In this case, the wavefunction falls in the right well. The central panel refers to the case in which the input tone is not applied. When at $t = t_1$  the potential is flipped in its double-well configuration, the wavefunction is split in half and we have an equal probability of measuring it in the left and right well.  (B) While falling in the well, the JDPD's wavefunction performs damped non-harmonic oscillations which leads to a gradual increase of the frequency }
    \label{fig:panel_wavefunction}
\end{figure}
The parameters introduced above provide E$_{L}$ = 743 GHz, E$_{J}$ = 2682 GHz and E$_C$ = 0.28 GHz, according to Eqs. \ref{eqn:potential_symm} and \ref{kinetic_energy}. It is worth noting that the system is dominated by the Josephson and inductive contributions, which lead the wavefunction to be very narrowed with respect to the quantum mechanical operator $ \varphi$ \cite{Clarke2008Jun}. The energies contributions lead to a frequency of $ \omega_0 / 2 \pi$ = 41 GHz when the potential is in the harmonic shape and $ \omega / 2\pi$ = 112 GHz when it is in the double well configuration. \newline
To model the dissipation in the presented device, the relaxation rate $\gamma$ has been extracted from numerical analysis performed in PSCAN2 \cite{pscan2}. This simulator allows to study of the classical dynamics of the circuit and includes a solid model to describe the Josephson junctions based on the \textit{Tunnel Junction Microscopic (TJM) } model  \cite{Werthamer,Ahmad2020Oct,Gulevich2019Feb,algorithms}. The parameters required to model the JJs have been determined in agreement with the typical fabrication values. In particular, we have considered I$_c$ = 5.4 $\mu$A, gap voltage V$_g$ = 2.6 mV, McCumber parameter \cite{barone} $\beta_C$ = 53, I$_c$ R$_N$/ V$_g$  ratio  = 0.8 and normal-to-subgap resistance ratio \cite{pscan2} R$_n$/R$_{sg}$ = 0.1. We have estimated $1/\gamma \simeq$ 7 ps, which is comparable to dissipation rates reported in the literature for similar standalone QFP devices \cite{1065574_damping,Takeuchi2014Nov_damping,Takeuchi2017Mar}. It is worth noting that in future design this value can be properly adjusted by changing the capacitance C or the effective resistance R of the junctions. We can also increase the $\beta$ factor, associated with the barrier height, to reduce the effect of dissipation on the phase trapping.\newline
Simulations have been carried out by considering a sinusoidal input tone with a frequency of  7 GHz  modulated by a Gaussian envelope with standard deviation of 100 ps. The input tone frequency sets the time scale for the flux-switch. Intuitively, the barrier height that separates the two wells should rise faster than the oscillations induced by the input tone, to prevent the redistribution of the probability between the two possible states.
In the case of our simulations, we assume that the duration of the flux-switch t$_{flip}$ is equal to 50 ps corresponding to a 20 GHz tone in the frequency domain. This value comes from the fact that our future idea is to provide the flux-switch signal with an integrated SFQ circuit and a 20 GHz clock has been proposed to be optimal in this type of superconducting device\cite{oleg1}. \newline
The amplitude of the input stimulus is expressed in terms of the wavefunction maximum elongation:
\begin{equation}
     \varphi_{0} = 2 \pi L I_{0} / \Phi_0 
\end{equation}
In the case of simulation in Fig. \ref{fig:panel_wavefunction}, we have chosen:
\begin{equation}
    \varphi_{0} = 2 \sigma
\end{equation}
where $\sigma$ is the JDPD's wavefunction standard deviation, according to Eq. \ref{eq:stnd_dev}. This value results to be optimal to achieve high fidelity, as shown in Fig. \ref{fig:num_analysis} B, where we will address  phase detection performed with other values of the stimulus's amplitude. \newline
The wavefunction time evolution is reported in Fig. \ref{fig:panel_wavefunction} A. The system is initialized in the harmonic state and the wavefunction is expected to collapse in the ground state of the potential, corresponding to the ``Ready" step in Fig. 2. At t = t$_0 $, the input signal starts to drive the wavefunction around the minimum. In the simulations, the sinusoidal waveform has a duration of five periods, which corresponds to 715 ps for a frequency of 7 GHz. Driven coherently by the external tone, the JDPD's wavefunction is described by a Gaussian wave-packet, in which the mean value $< \varphi(t)>$ evolves as:
\begin{equation}
\label{eq:drive}
    <  \varphi  >  = \varphi_{0} sin(\omega t + \theta_0 )
\end{equation}
where $\omega / 2 \pi  = 7 \; GHz $ and $\theta_0$ is the displacement of the sinusoidal input signal.
As discussed in the section ``Principle of Operation", the wavefunction standard deviation depends on device parameters and it is constant during the driven evolution. According to Eq. \eqref{eq:stnd_dev}, we have estimated $\sigma$ = 0.31 rad considering the values of L and C indicated above.
Discrimination between the two states is made at t=t$_1$ when the flux-switch is applied. In simulations, t$_1$ is chosen to center the flux-switch with respect to the beginning  and the end of the input tone. The wavefunction collapses in either the left $|L>$ or right $|R>$ state depending on the initial displacement $\theta_r$.
For instance, the left panel of Fig. \ref{fig:panel_wavefunction} shows the wavefunction evolution considering an initial $\theta_r = 0$, which leads the wavefunction to fall in the left well after the flux-switch pulse. In the right part of Fig. \ref{fig:panel_wavefunction} it is represented the opposite situation, corresponding to $\theta_r = \pi$. The central panel part to the case in which the input tone is not applied. When at t=t$_1$  the potential is flipped in its double-well configuration, the wavefunction is split in half and we have an equal probability of measuring it in the left and right wells. \newline
While collapsing in the well, one may ask which is the induced backaction on the input signal's source. In this transient, the JDPD can be modeled as an effective voltage source $V_j$ that can generate a disturbing signal on the rest of the circuit \cite{Opremcak}. According to the ac Josephson relation \cite{barone,tafuri}, the produced backaction signal is proportional to the time derivative of $<\dot \varphi(t)>$. After the flux-switches the JDPD's waveform performs a damped non-harmonic motion, where the oscillations frequency gradually increases \cite{backaction_1,backaction_2}. Since phase detection does not require the JDPD to be in resonance with the input signal source, one can design the detector in such a manner that these oscillations are outside the source's absorbing spectrum preventing to induce backaction on the system. For example, considering the selected parameters for simulations, these oscillations have frequencies above 90 GHz, as shown in Fig. \ref{fig:panel_wavefunction} B. They stabilize around 112 GHz,  which corresponds to the frequency of the two wells when the potential is set in the flipped configuration (i.e $\phi_+ = \pi$).\newline
The sequence is completed at t=t$_2$ when the position of the phase particle can be ``Sensed" as the two possible outcomes of a measurement of the ${\varphi}$ operator will have opposite values, which correspond to opposite signs for the current flowing through the central inductor L:
\begin{equation}
    <I> = \frac{\Phi_0}{2\pi L} <\varphi>.
\end{equation}
\begin{figure}
\centering
\includegraphics{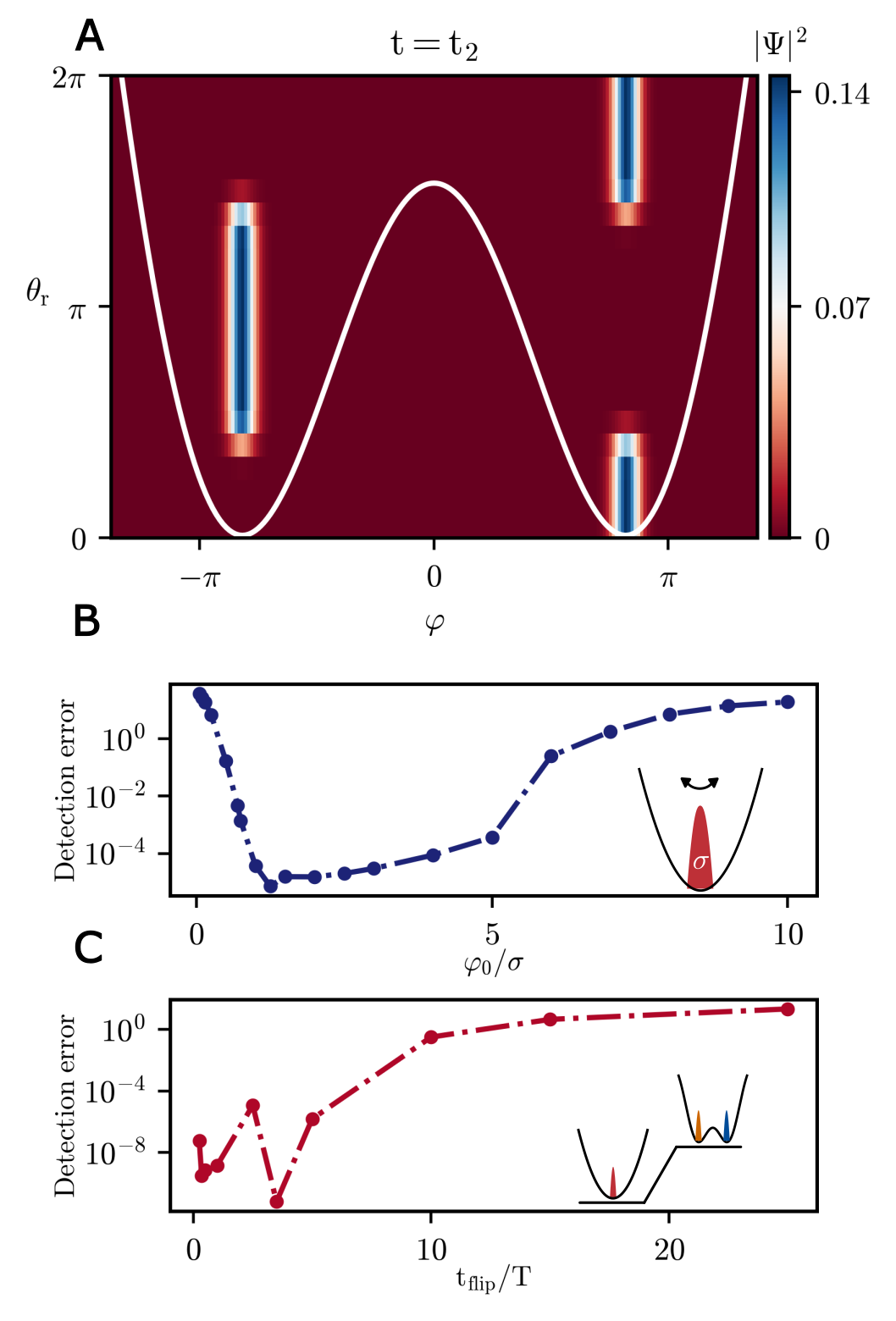}
\caption{(A) JDPD's probably distribution $|\Psi(t)|^2(t)$ at t=t$_2$ as a  function of the initial displacement $\theta_r$ of the input tone. \newline (B) Detection error as a function of the wavefunction's maximum elongation $\varphi_{0}$ during the driven harmonic motion. $\varphi_{0}$ is expressed in unit of the wavefunction's rms $\sigma$. \newline (C) Detection error as a function of flux-switch rise time t$_{flip}$ normalized  in terms of input signal's period T.}
\label{fig:num_analysis}
\end{figure}
Fig. \ref{fig:num_analysis} A exhibits the wavefunction's probability density $|\Psi|^2$ at t=t$_2$ for several values of the input tone's displacement $\theta_r$. According to these simulations, a fidelity close to 1 is in principle achievable; however, this estimation depends on several parameters.
Fig. \ref{fig:num_analysis} B shows the detection error as a function of the wavefunction's maximum elongation $\varphi_{0}$ in units of the wavefunction's standard deviation $\sigma$:
\begin{equation}
    \varphi_{0}  \equiv n \sigma
\end{equation}
Fidelity is maximized when $ n \in [1,6]$. Values below this range correspond to small oscillations around $\varphi = 0$. As a consequence, when the flux-switch is applied, the wavefunction is split between the two  wells leading to a reduction in measured fidelity. When $n > 6$, the barrier height is not large enough to confine the  wavefunction  and this leads to a redistribution of probability.\newline
Another factor that limits fidelity is the flux-switch duration t$_{flip}$. According to simulations in Fig. \ref{fig:num_analysis} C, a fidelity close to $1$ is achievable when the t$_{flip} <$  10 T, where  T = 2$\pi/\omega$ and $\omega / 2 \pi$ = 7 GHz is the simulated frequency for the input tone. In fact, a slow rise time for the barrier height induces a redistribution of probability between the two wells at each input tone's period. Considering the typical frequencies for superconducting resonators, optimal values for t$_{flip}$  are in the order of hundreds of ps. We want to point out that the possibility to manipulate diabatically the potential shape  has been already exploited both experimentally and theoretically in  Refs. \cite{Poletto_2009,Castellano_2010,Chiarello_2012}. These papers demonstrated coherent oscillations of a tunable superconducting flux qubit by manipulating its energy potential with nanosecond-long pulses of magnetic flux. Given the resemblance between the JDPD and the device reported in Refs. \cite{Poletto_2009,Castellano_2010,Chiarello_2012}, we can argue that fast potential manipulation can be performed also in the case of the presented detector. \newline
A possible approach to generate pulses with hundreds of ps rise time involves the use of a dedicated SFQ flux generator, which can operate with a clock of tens of GHz \cite{oleg1,mcdermot_sfq}.
Another benefit of the use of high-speed SFQ circuitry lies in the possibility of performing multiple measurements on the same cycle, thus allowing averaging to reduce the noise contribution and improve fidelity.
This is shown in Fig. \ref{fig:multiple_flips} where the measurement sequence with multiple flux-flips has been simulated in QuTip. \newline
A more detailed discussion about the integration of the JDPD with an SFQ-based platform can be found in the Section ``Outlook”.
\begin{figure}
    \centering
    \includegraphics{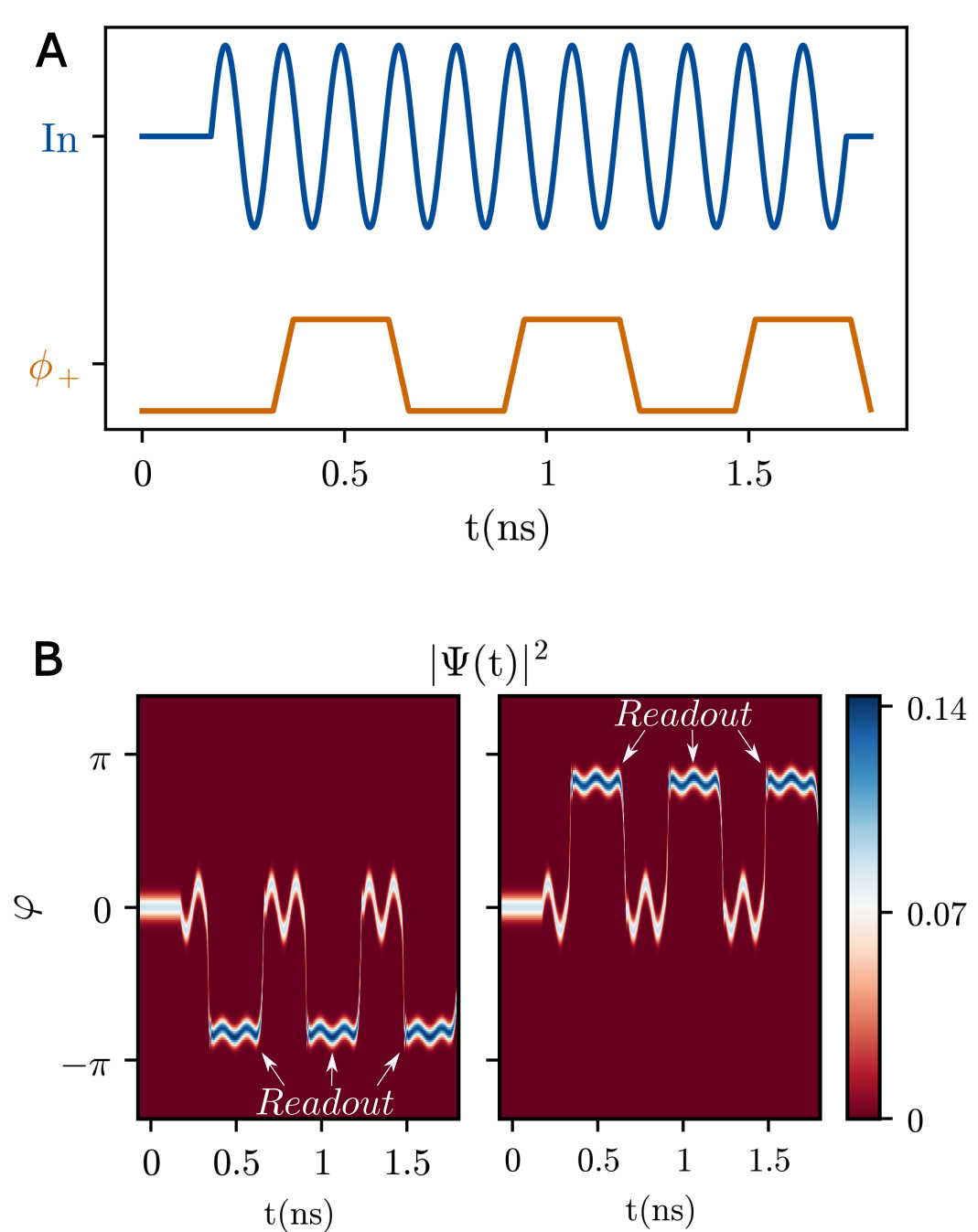}
    \caption{Multiple phase detections are performed on the same sinusoidal input tone. (A) Time diagram of the applied signals. The flux-switch curve is orange while the sinusoidal input tone is blue. (B) The time sequence reported in A has been simulated in QuTip and the $|\Psi(t)|^2$ is shown. In particular, in the left panel, the sinusoidal input tone is initialized with displacement $\theta_0$=0 while in the right one $\theta_0=\pi$}
    \label{fig:multiple_flips}
\end{figure}

\section{Device design and experimental setup}
\begin{figure}
    \centering
    \includegraphics{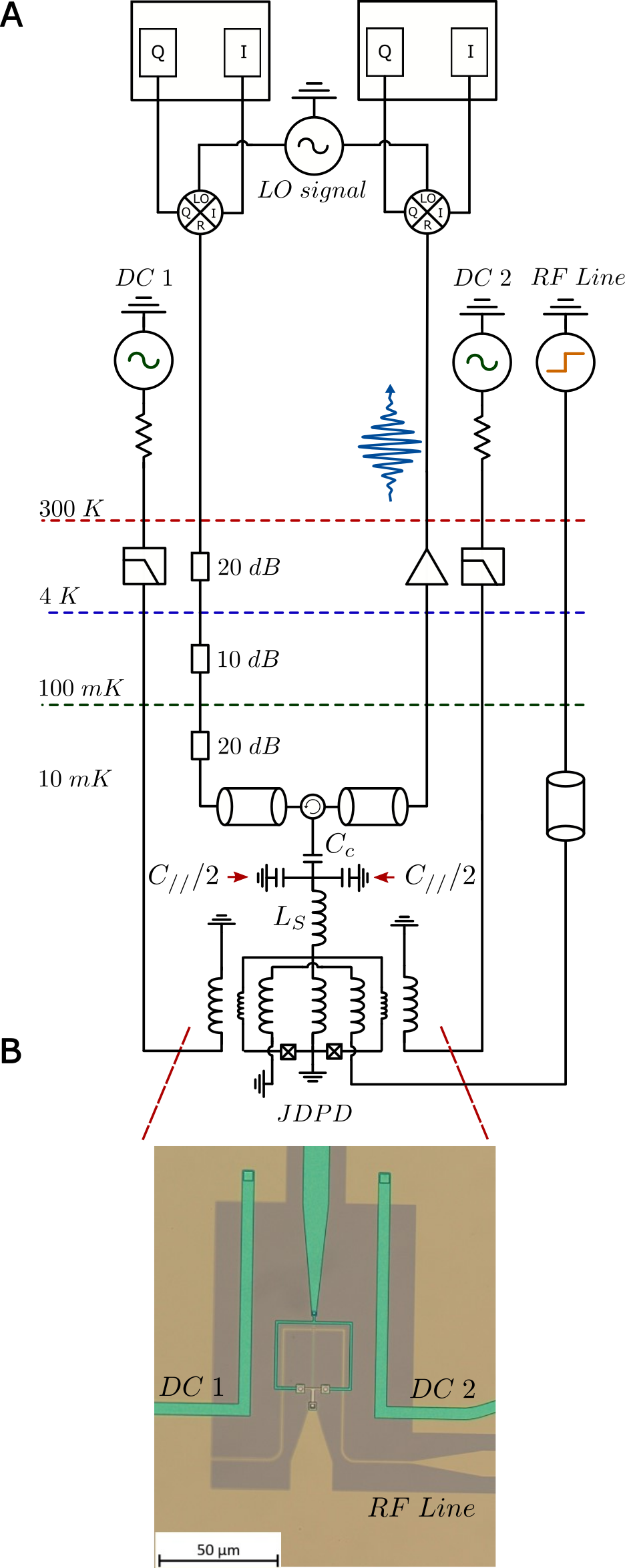}
    \caption{ (A) Measurement setup for the characterization of the JDPD. The device has been measured in reflection through the standard heterodyne detection technique. The JDPD acts like a flux tunable inductance which changes the resonance frequency as a function of the applied fluxes $\phi_+$. These two fluxes act differently on the JDPD's potential producing a flip or controlling the asymmetries with respect to $\varphi = 0$. The magnetic flux is provided by an rf flux line and two independent dc flux lines connected to two AWG at room temperature.  (B) Optical microscope image of one of the realized devices .\newline}
    \label{fig:panel_exp_setup}
\end{figure}

The JDPD has been realized in collaboration with SEEQC \cite{Seeqc} using their multilayer fabrication process. The device micrograph is reported in Fig. \ref{fig:panel_exp_setup} B. The JDPD central inductor L has been realized by a 40-$\mu$m-long and 1.4-$\mu$m-wide wire fabricated using a high-kinetic-inductance layer in NbN, with a total value of L = 220 pH. The rest of the circuit is made with conventional  layers in Nb, characterized by smaller values of sheet inductance. The Josephson junctions are realized with a trilayer stack of Nb/AlOx/Nb with 1 kA cm$^{-2}$ critical current density. In particular, we have chosen Josephson Junctions with an I$_c$ of 5.4 $\mu$A,  corresponding to 2 $\beta_L = 7 $. As discussed in the Section ``Device model", this value is large enough to produce a double wells potential when the device is flux-switched. \newline
The JDPD is equipped with two dc lines, $DC1$ and $DC2$, and an $RF \;line $, that allow us to have precise control of the magnetic fluxes threading the device's loops. The $RF \;line $ is designed to pass under the JDPD main loop, as shown in Fig. \ref{fig:panel_exp_setup} A and  Fig. \ref{fig:panel_exp_setup} B. Considering its geometry, the $RF \;line $ forces the current to circulate around the JDPD main loop without flowing through the central inductor L, which contributes only to the common flux $\phi_+$.
The flux $\phi_+$ allows us to set and reset the JDPD, changing the potential from a harmonic configuration to a double well one and vice-versa. According to simulations reported in the Section ``Numerical Analysis", these operations should be performed diabatically to enhance the detection fidelity and the protocol speed. For this reason, we have connected the $RF \;line$ below the JDPD loop to a conventional 50 $ \Omega$ Coplanar Waveguide (CPW) to preserve as much as possible the shape of the flux-switch. The CPW is made of an 11-$\mu$m-wide superconducting core in $N_b$ and it has a gap of 7 $\mu$m to ground.\newline
The lines  $DC1$ and $DC2$ provide individually a contribution in both the JDPD meshes. To be more clear, defining I$_{DC1}$ and  I$_{DC2}$ as the currents flowing respectively in $DC1$ and $DC2$, we have:
\begin{equation}
\begin{gathered}
\phi_1 \frac{\Phi_0}{2\pi} = M_{1,1} I_{DC1} \; + M_{1,2} I_{DC2}  \\
\phi_2 \frac{\Phi_0}{2\pi} =  M_{2,1} I_{DC1} \; + M_{2,2} I_{DC2} 
\end{gathered}
\end{equation}
where M$_{i,j}$ is the mutual inductance between the dc line $j$ and the mesh $i$. As shown in Fig. \ref{fig:panel_exp_setup} B, the two dc lines are placed on opposite sides with respect to the central inductor L, which indicates that the matrix of coefficients M$_{i,j}$ is:
\begin{equation}
\begin{gathered}
    M_{1,1} =  - M_{2,2} = M_{dir} \\
    M_{1,2} =  - M_{2,1} = M_{opp}
\end{gathered}
\label{eq_mutula}
\end{equation}
According to Eqs. \ref{eq_mutula}, any combination of $\phi_+$ and $\phi_-$ can be generated by properly biasing the two dc lines. Cases of particular interest occur when $\phi_+ = 0$, which implies $I_{DC1} = I_{DC2} $:
\begin{equation}
\begin{gathered}
0 = \frac{\phi_1 + \phi_2}{2} \frac{\Phi_0}{2\pi} =\frac{ M_{dir}}{2} \bigg(I_{DC1} - I_{DC2} \bigg) +\\ \frac{ M_{opp}}{2} \bigg(I_{DC1} - I_{DC2}\bigg) \\
\implies I_{DC1} =  I_{DC2}
\end{gathered}
\end{equation}
and when $\phi_- = 0$, leading to $I_{DC1} = - I_{DC2} $: :
\begin{equation}
\begin{gathered}
0 = \frac{\phi_1 - \phi_2}{2} \frac{\Phi_0}{2\pi} =\frac{ M_{dir}}{2} \bigg(I_{DC1} + I_{DC2} \bigg) + \\ \frac{ M_{opp}}{2} \bigg(I_{DC1} + I_{DC2} \bigg) \\
\implies I_{DC1} = -  I_{DC2}
\end{gathered}
\end{equation}
The two dc lines are employed to control the asymmetries of the system and to set the JDPD working point. The dc lines are realized with a 5-$\mu$m-wide wire in $Nb$. Both rf and dc lines are connected at room temperature to an AWG with a 1 GS/s sampling rate. \newline
The JDPD dynamics is investigated through reflection measurements by adopting a typical heterodyne detection setup as indicated in Fig. \ref{fig:panel_exp_setup} A. To shift the resonance frequency in the measurable range [4 GHz,8 GHz], a lumped $LC$ resonator has been coupled to the device. The circuit comprises an inductor L$_s$ in series with the JDPD and a parallel plate capacitor C$_{//}$ close to the ground. A coupling capacitor C$_{c}$ connects this resonator to the external experimental apparatus, as shown in Fig.\ref{fig:panel_exp_setup} A. At linear order of approximation, the JDPD works as a lumped variable inductance L$_\mathrm{JDPD}$ which depends on $\phi_+$ and $\phi_-$: 
\begin{equation}
    L_\mathrm{JDPD} (\phi_+,\phi_-) = \left(\frac{\Phi_0}{2\pi}\right)^2\frac{1}{\left.\frac{d^2 U(\varphi) (\phi_+,\phi_-)}{d\varphi^2}\right|_{\varphi_{min}}}
\end{equation}
where $\varphi_{min}$ is the potential energy minimum where the phase particle is trapped. In this way, the system is characterised by a resonance frequency:
\begin{equation}
\label{eq:resonance frequency}
    \omega / 2 \pi = \frac{1}{\sqrt{\left ( C_c + C_{//}  \right ) \left ( L_{JDPD} + L_s  \right )}}
\end{equation}
which is directly linked to the JDPD state.\newline
The values of L$_s$, C$_{c}$  and C$_{//}$ have been carefully chosen to deliver signals out of resonance through the JDPD input node, as requested by the phase detection technique while preserving the resonance visibility. These constraints lead us to consider L$_s$ = 300 pH, C$_{c}$ = 100 fF and C$_{//}$ = 1.4 pF, determined in agreement with numerical simulations.
The capacitors $C_{c}$ and $C_{//}$ have been realized in a parallel plate configuration between two $Nb$ layers  separated by a dielectric part in $SiO_2$, characterized by a specific capacitance of 0.44 fF/ $\mu$m$^2$. This choice is obligatory since it's not possible to achieve large values of capacitance with interdigitated capacitors, as required in the present circuit. The drawback is that the typology of  capacitors leads to higher values of the circuit internal losses \cite{Opremcak,Opremcak2} and they are more prone to manufacture parameters spread.  \newline
The inductor L$_s$ is realized using the high kinetic inductance layer in $NbN$, which brings to a large value of L$_s$ in a relatively small portion of space.
The measurements were performed at about 10 mK using a dry dilution fridge.
\newline
\section{Experimental results}
Device characterization starts by performing spectroscopy versus fluxes $\phi_+$ and $\phi_-$. Comparing the resulting map with simulations, we can determine the locations of the bias points corresponding to the potential energy configurations that the JDPD can assume during the protocol. We have measured a resonance frequency of 6.6 GHz when the JDPD is set in the state $\phi_+$ = 0, 6.4 GHz when the JDPD is in the harmonic configuration and 6.588 GHz $\phi_+$ = $\pi$.
In the harmonic state, we have estimated a resonator quality factor $Q_{int} \simeq 70$,  corresponding to a decay rate of $\kappa = 1/$(0.22$ ns$).
\newline
Phase detection requires the determination of the symmetry point $\theta_{symm}$ concerning the flux $\phi_-$. As suggested by the name, $\theta_{symm}$ is the value of $\phi_-$ that leads to an equal splitting of the wavefunction in the two wells when a flux-switch is applied in the absence of any external input signal. Theoretically, $\theta_{symm} $ should correspond to $\phi_- = 0$; however, some factors, such as asymmetric junctions or fabrication spread, could lead to $\theta_{symm} \neq 0 $ as discussed in the Appendix. The determination of $\theta_{symm}$ is crucial in order to achieve state-independent fidelity.
\begin{figure}[!]
    \centering
    \includegraphics[trim=0.7 2 2 2,clip]{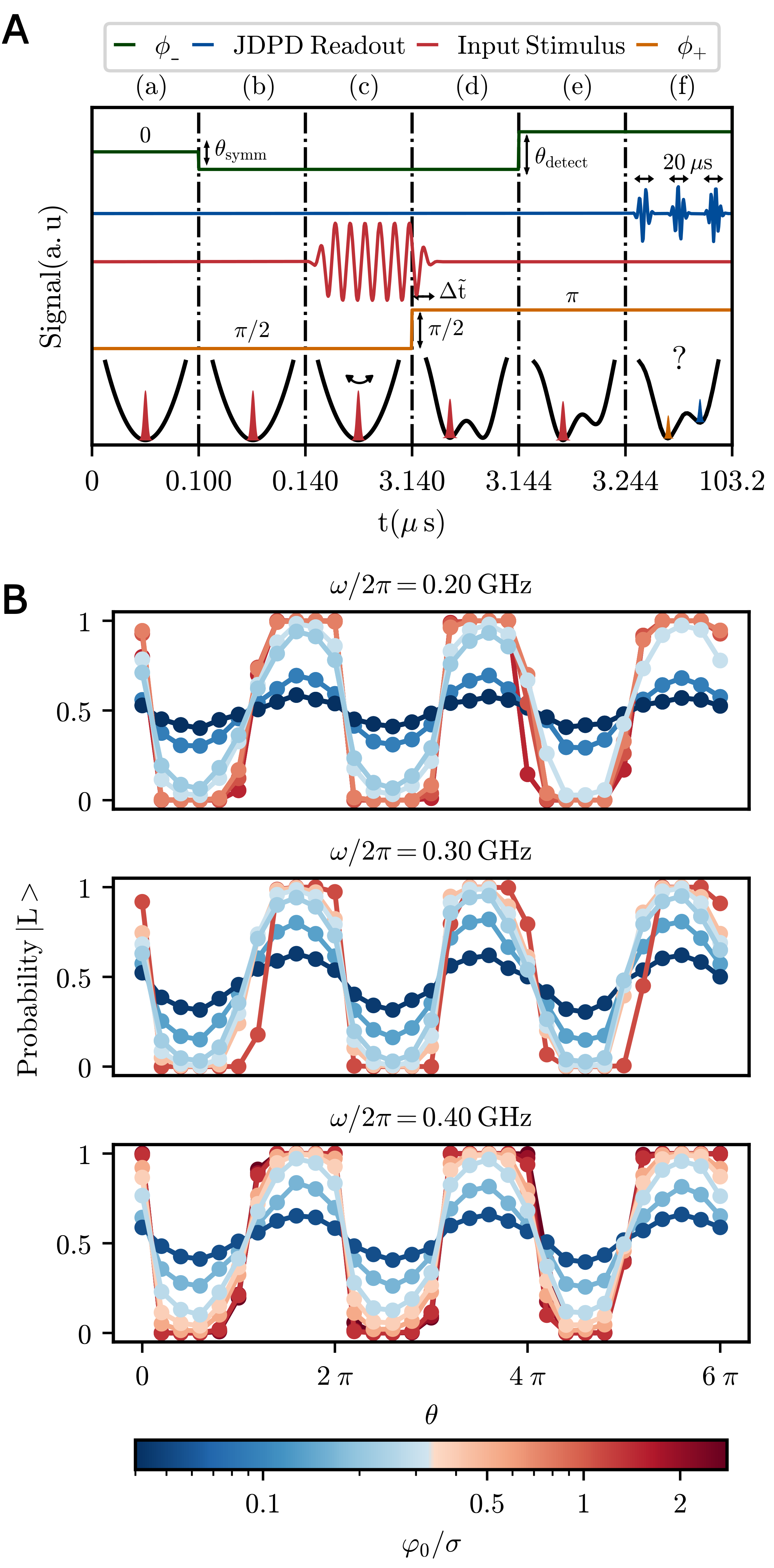}
    \caption{(A) Time sequence performed experimentally to demonstrate phase detection, as described in details in the text. (B) Phase detection has been performed for different values of input frequency and amplitude expressed in terms of $\varphi_0/\sigma$, as introduced in the Section "Numerical Analysis". Results are compatible with the theoretical expectations and demonstrate the feasibility of the JDPD approach.}
    \label{fig:dc_detect}
\end{figure}
Once we have calibrated the system, we have evaluated experimentally the JDPD's capability to work as a phase detector, according to the time sequence reported in Fig. \ref{fig:dc_detect} A. \newline
Phase detection begins by preparing the system in the harmonic configuration. We wait a cooldown time of 100 ns to ensure that the wavefunction collapses in the minimum of the potential energy. This step forces the device to ``Reset" in the ground state  and makes it ready to be flipped in the double well configuration. At t = 100 ns, we apply $\phi_{-} = \theta_{symm}$, determined during the calibration. In step (c) of Fig. \ref{fig:dc_detect} A, the JDPD is driven by an input stimulus for a total duration of  3 $\mu$s. The application of this tone makes the wavefunction oscillate coherently around the potential minimum, as expected from Eq. \ref{eq:drive}. The potential is flipped in the double well configuration at 3.140 $\mu$s, as indicated in Fig. \ref{fig:dc_detect} (A). The flux-switch is provided diabatically, with a rise time of 1 ns corresponding to the maximum achievable by AWGs. The effect of 
 the potential flipping makes the wavefunction collapse in the left $|L>$ or $|R>$ state, depending on its position with respect to $\varphi = 0$ when the flux-switch is applied. To ensure the overlap between the pulses, we retarded the end of the stimulus by $\Delta \tilde t$ = 4 ns with respect to the beginning  of the flux-switch. \newline
 The readout of the JDPD state is accomplished in steps (e) and (f). In the actual chip design, there is no possibility of measuring the superconducting current that passes through the inductor L and the JDPD's readout is performed by using the spectroscopy measurement. However, in the symmetric condition when left $|L>$ and $|R>$ states are equiprobable, the two wells have the same resonance frequency and it's tricky to distinguish them. \newline
 To overcome this difficulty, we can unbalance a little bit the frequency of the two states by applying $\phi_- = \theta_{detect}$. Reasonably, $\theta_{detect}$  not to perturb too much the system too much. In the case of measurement, we have chosen  $\theta_{detect} =$ 0.2 rad, which has proven experimentally to be a good value for the observability of the $|L>$ and $|R>$ states. The JDPD is, then, measured in reflection at 5 probing frequencies; each readout pulse has a lenght of 20 $\mu$s corresponding to 100 $\mu$s of total duration for step (f). \newline 
 The sequence described above is repeated 5000 times to make statistics on the trapping probability. As illustrated in Fig. \ref{fig:dc_detect} B, the detection sequence has a total duration of several $\mu$s, primarily due to step (f) where we measure in reflection the JDPD state. However, in future layouts, we will probe the device state by digitalizing the sign of the current flowing through the inductor L. This task can be efficiently performed by an SFQ current comparator, which can operate at tens of GHz clock speed \cite{oleg1,x3} bringing important benefits to reduce the duration of step (f).  Furthermore, step (e), in which the device frequency is unbalanced, is no longer required and $\phi_- = \theta_{symm}$ can be set at the beginning of the calibration and left untouched during the sequence. Since detection will not be performed by measuring the JDPD resonance frequency, we can opportunely modify the parameters of the resonator coupled to the device to minimize the input stimulus's ringing time and, consequently, the duration of step (C). From these arguments, we conclude that the timing reported in Fig. \ref{fig:dc_detect} B can be extremely reduced to a few nanoseconds time scale, as theoretically predicted in Fig. \ref{fig:panel_wavefunction} A. \newline
 Experimental outcomes are reported in Fig. \ref{fig:dc_detect} B. The detection protocol has been performed by changing the stimulus initial phase $\theta$ in the range $\theta \in [0,6\pi]$ for different values of input frequency up to 400 MHz, which is the maximum tone frequency that our instruments can generate with a controlled input phase. We have also swept over the amplitudes $A_m$, expressed in terms of $\varphi_0/\sigma$. \newline 
Results show that JDPD is sensitive to the applied input tone. In particular, when the amplitude $A_m$  is large enough, as in the case of $\varphi_0/\sigma > $ 0.55, the device is able to perform the digitalization of stimulus phase $\theta$, as this information is mapped into the wavefunction collapsing probability in either the two states $|R>$ and $|L>$. The grey zone reasonably decreases with the amplitude $A_m$ until, in the case of small input amplitudes, the wavefunction is no longer capable of reaching the $|L>$ and $|R>$ states with probability approaching 1. In this case, the interpolation of points, the  $Probability \; |L>$ vs $\theta$ exhibits a sinusoidal shape compatible with the input tone profile. When $A_m$ increases, the effect of digitalization leads to a "clipping" of the interpolation shape, which is more evident for the red points in Fig. \ref{fig:dc_detect} B. This behavior is compatible with the theoretical prediction reported in Fig. \ref{fig:num_analysis} A, where the wavefunction probability distribution $|\Psi(t)|^2$ is reported as a function of the initial displacement $\theta$. \newline
\begin{figure}
    \centering
\includegraphics{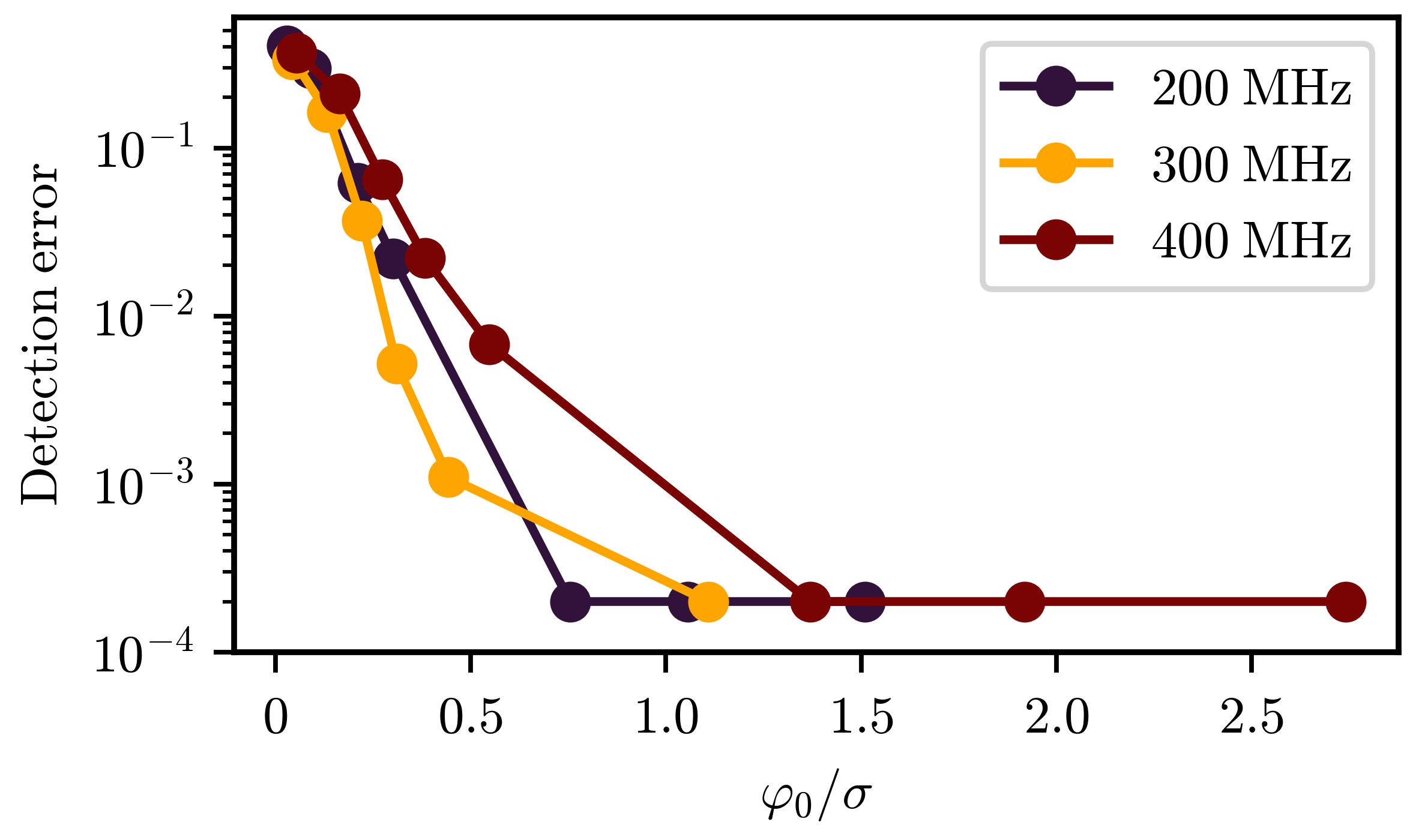}
    \caption{Detection error vs the input tone amplitude, expressed in terms of $\varphi_0/\sigma$. Results are compatible with theoretical prediction reported in Fig. \ref{fig:num_analysis} B.}
    \label{fig:readout_error}
\end{figure}
 Fig. \ref{fig:readout_error} shows the dependence of the detection error on the stimulus amplitude $A_m$. According to experimental outcomes, phase detection can be achieved with fidelity of  99.98 $\%$, in  a similar way compared to numerical analysis in Fig. \ref{fig:num_analysis} B. 
\section{Outlook}
\begin{figure}[ht!]
    \centering
    \includegraphics{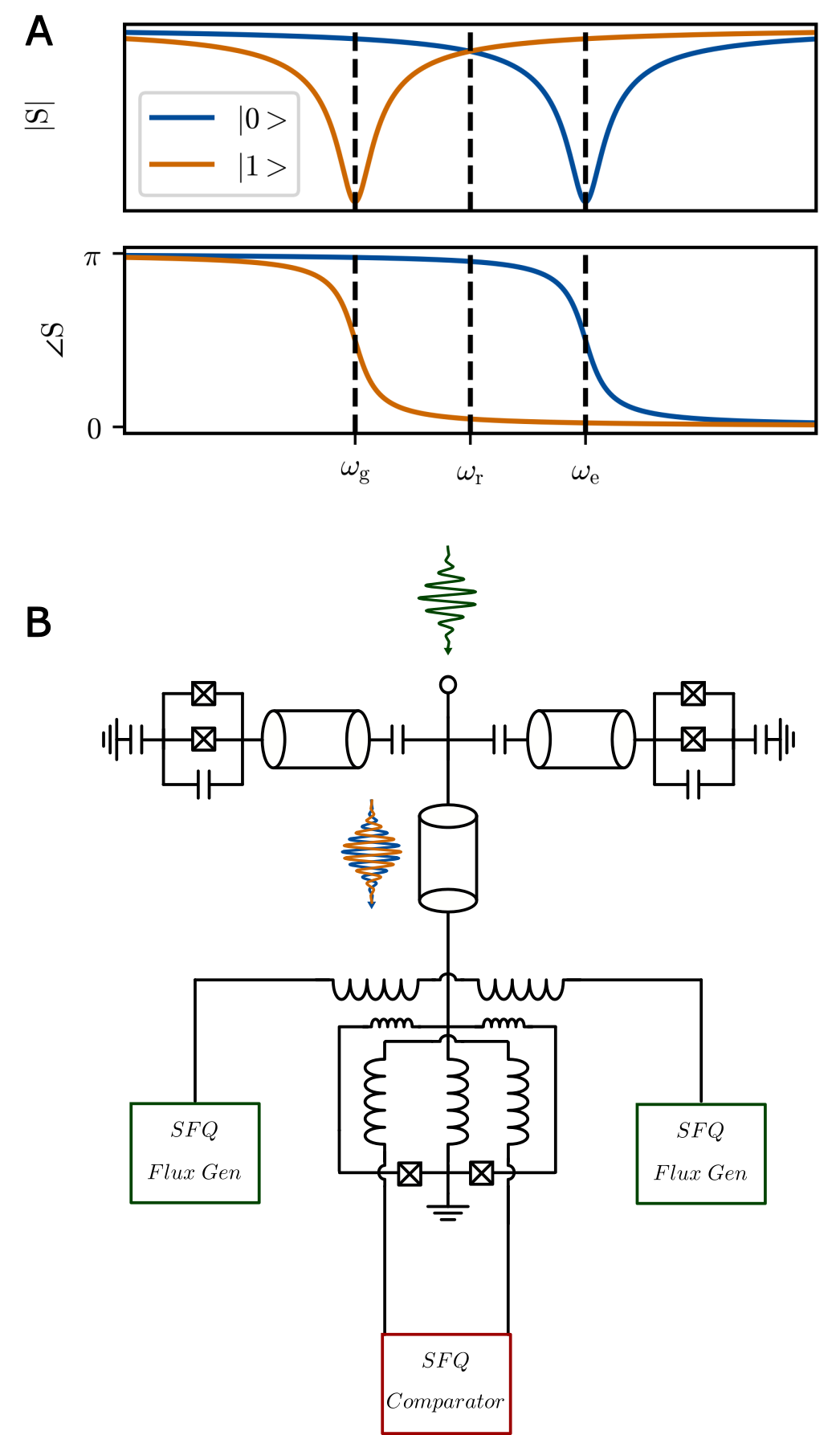}
    \caption{(A) $S$ parameters measured in a typical superconducting qubit readout scheme. A shift of the resonance frequency is observed depending on the qubit's state, which is visible both in amplitude $|S|$ and phase $\angle S$ and. To exploit the JDPD detection technique, we need to maximize the separation between the state $|0>$ and $|1>$ in phase. This operation can be done by probing the cavity at the resonator bare frequency,  i.e $\omega = \omega_r$. (B)  Possible experimental setup to measure the qubit's state involving the JDPD. In this sketch, two resonator-qubit systems are capacitively coupled to a feedline connected inductively to a JDPD device. The detector's states are tuned by an SFQ flux generator and the measurement output is digitalized by using an SFQ comparator.}
    \label{fig:qubit_setup}
\end{figure}
In this work, we have illustrated a phase detection technique based on a Josephson Digital Phase Detector (JDPD). Simulations and Experimental outcomes show that digitalization can reach close-to-1 fidelities in a nanosecond timescale with negligible backaction on the source. These characteristics make the JDPD approach particularly suitable for the readout  of superconducting qubits. \newline
Typical schemes for superconducting qubit measurement exploits high
$Q$ resonators whose bare resonant frequency $\omega_R$ is shifted
by a factor $\chi$, depending on the state of a dispersively coupled qubit. Discrimination can be performed by probing the resonator with microwave tones and measuring the response in phase and amplitude as reported in Fig. \ref{fig:qubit_setup} A.
A similar approach can be pursued in the case of the presented detection technique, as schematized in Fig. \ref{fig:qubit_setup} B. We can probe the cavity by sending a tone at $\omega = \omega_r$ to be maximally sensitive to the phase response of the dressed resonator. Subsequently, the output signal can be sent to an inductively coupled JDPD set to be in the ``Ready" state. The JDPD will behave almost as a purely inductive element at input frequency $\omega_r$, so $\varphi(t)$ will be phase-locked to the current injected by the readout tone. \newline
The operations needed to drive and measure the JDPD can be performed by an SFQ circuitry \cite{x1,x2}. These devices store the bits of information using propagating fluxons, as picosecond voltage pulses whose time integral equals the superconducting flux quantum $\Phi_0 = h/2e$. Thanks to their very low power dissipation, energy-efficient SFQ (ERSFQ,eSFQ) circuits \cite{5682046_esfq,5688194_esfq2,Mukhanov1987_sfq3} could be safely located contiguously to the quantum chips and they offer the possibility to operate at very high speeds (tens of gigahertz clock) with small jitter \cite{qubit_control_mcdermott}. Therefore, an SFQ circuitry can grant the required phase locking between the input pulse and flip pulse making the detection possible with fidelities close to 1. \newline
To realize the discussed system, we are currently testing an optimized JDPD and designing a complete circuit which includes an SFQ flux generator\cite{qubit_control_mcdermott}, to rapidly flux-switch the potential, and an SFQ comparator \cite{caleb1,1211768_83,4277641_86,PhysRevLett.89.217004_82}, to measure the JDPD state.

\section{Conclusions}
In conclusion, we have reported a phase detection technique based on Josephson Digital Phase Detector.  This device is based on a QFP and has a tunable potential which can be diabatically switched from a single-minimum-potential to a double-minima configuration.  If a coherent signal is applied, the system wavefunction collapses in either of its two degenerate wells depending on the phase sign of the input tone. This information is therefore mapped in the current direction flowing in the JDPD central linear inductor. The proposed measurement scheme is performed directly on chip at the millikelvin stage of a dilution refrigerator; furthermore, it eliminates the need for microwave output lines which brings significant technical and economic challenges to system scaling. 
The JDPD works far detuned with respect to the qubit, thus reducing the effects of backaction. The basic concepts behind this detection scheme have been experimentally verified. The capability to work as a phase detector has also been demonstrated up to 400 MHz tone with a remarkable agreement between the experimental outcomes and simulations. As a future perspective, we have discussed a possible implementation of this device to readout the state of a superconducting qubit, which can be accomplished by properly adjusting the JDPD design parameters. This detector is also well suited to work with a scalable SFQ architecture, which can be employed to drive and measure the JDPD state. Therefore, we envision the JDPD as part of a more complex architecture, in which the classical qubit control, measurement, and data processing are performed by a classical SFQ processor. This approach can be a valid solution to enhance the scalability of superconducting quantum systems \cite{oleg1,mcdermot_sfq}, which remains a big engineering challenge to realize practical error-corrected quantum computers \cite{PhysRevA.86.032324_surface}.
\begin{acknowledgments}{
The authors would like to thank D. Yohannes, J. Vivalda, M. Renzullo and A. Chambal-Jacobs for the fabrication of the samples and A. Kirichenko, A. Salim and R. Albu for advice in design.
\newline
The work was supported by the project ``On-chip signal generation for  superconducting Quantum Processors (SFQ4QPU)", in the frame of Eurostars CoD 15 Call 2021; the project ``SQUAD—On-chip control and  advanced read-out for superconducting qubit arrays (SQUAD)” in the  frame of Programme STAR Plus, financially supported by UniNA and  Compagnia di San Paolo; and the project ``Superconducting quantum-classical  linked computing systems (SuperLink)", in the frame of QuantERA2  ERANET COFUND in Quantum Technologies. The research activities were also supported by the PNRR MUR project PE0000023-NQSTI and the PNRR MUR project CN\_00000013 - ICSC.}
\end{acknowledgments}
\bibliography{apssamp}
\newpage

\begin{section}{Appendix}
A perfectly symmetrical JDPD is a fundamental requirement to obtain state-independent fidelity and guarantee the working condition of the detector. \newline
However, in a real device, the JDPD will likely deviate from being perfectly symmetric due to various factors  but mainly by parameter spread of the fabrication process and different flux trapped in the two JDPD branches.\newline
\begin{figure*}
    \centering
    \includegraphics{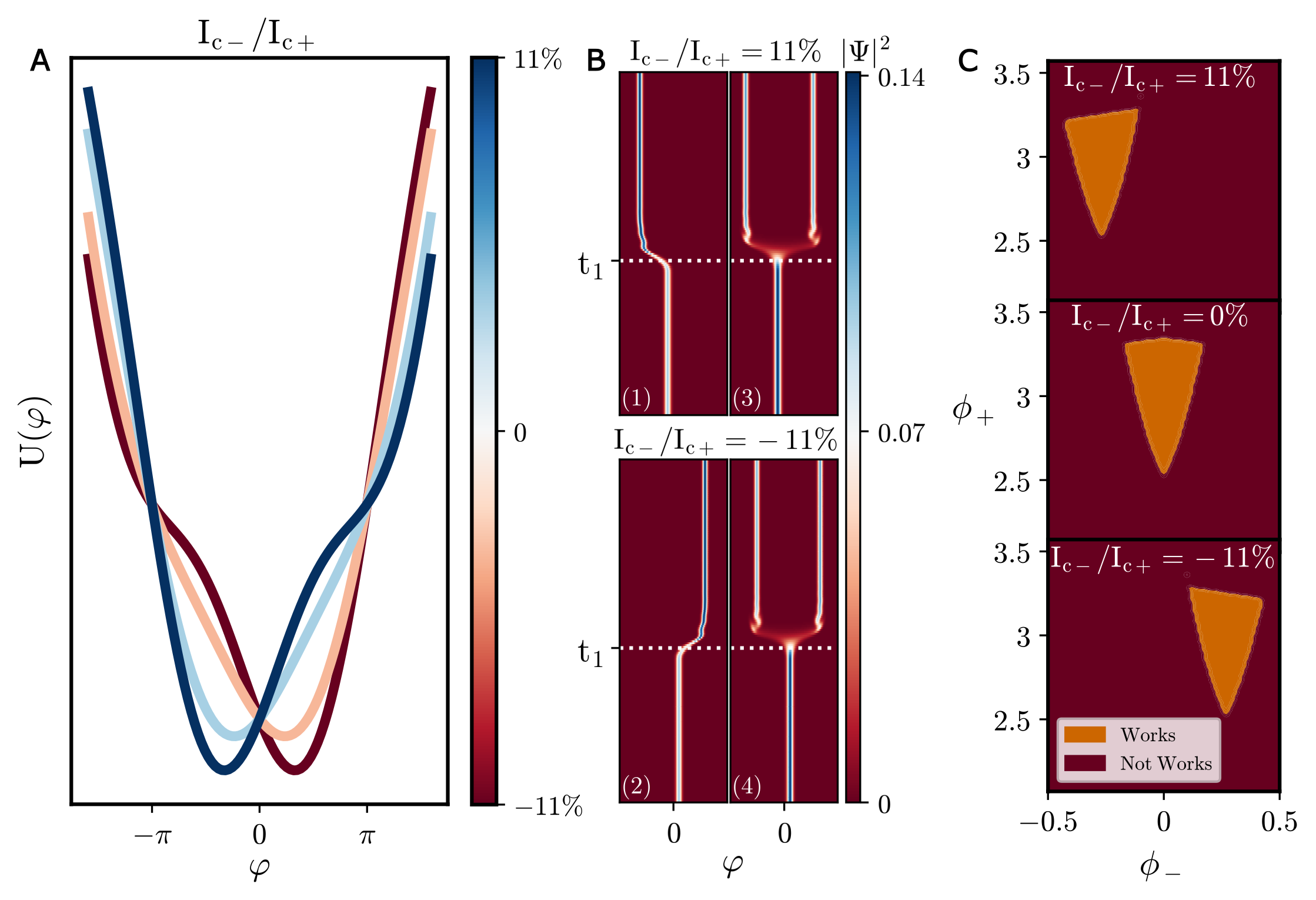}
    \caption{(A) JDPD's potential shape when $\phi_+ = \pi / 2$ and $\phi_- = 0$ for several values of the ratio I$_{c+}$/I$_{c-}$. The presence of asymmetries leads to a shift in the potential minimum and the wavefunction acquires an offset with respect to $\varphi = 0$. (B)(1-2) In the absence of external input, an applied flux-switch makes the wavefunction collapse in the two wells with different probabilities. In this QuTip simulation performed with I$_{c+}$ = 9 $\mu$A and I$_{c+}$/I$_{c-}$ = $\pm$ 11 $\%$, we have 99.999$\%$ of the probability of  reaching the left (1) and right well (2). (3-4) However, state-independent fidelity can be achieved by adjusting the value of  $\phi_- $ to  $\phi_- \simeq 0.46$ rad. (C) The numerical analysis performed with QuTip is supported by classical simulations made in PSCAN2 \cite{pscan2}. Sweeping on $\phi_{+}$ and $\phi_{-}$, we have determined regions in which the JDPD is capable of performing correctly phase detection. Simulations have been carried out considering the same parameters adopted for the QuTip's ones, i.e  I$_{c+}$ = 9$\mu$A and I$_{c+}$/I$_{c-}$ = $\pm$ 11 $\%$. In particular, the top figure  refers to the case in which I$_{c+}$/I$_{c-}$ = +11 $\%$, the central one corresponds to  perfect symmetrical JDPD and the one in the bottom panel shows the condition when I$_{c+}$/I$_{c-}$ = -11 $\%$. }
    \label{fig:fig_asymm}
\end{figure*}

In the case of an asymmetrical device, the potential energy becomes:
\begin{widetext}
\begin{equation}
    U(\varphi) = \frac{1}{2L} \biggl[ \biggl ( \frac{\Phi_0}{2 \pi  }\biggl)^2 \varphi^2 - \frac{\Phi_0}{2 \pi} \biggl ( I_{c+} \; cos( \phi_+ ) \;cos(\varphi \; + \; \phi_-) -  I_{c-} sin( \phi_+ ) \; cos(\varphi \; + \; \phi_-)\biggl ) \biggl ]
\end{equation}
\end{widetext}
where the variables $I_{c+}$ and $I_{c-}$ have been introduced:
\begin{equation*}
\begin{cases}
I_{c+} = I_{c1} + I_{c2} \\
I_{c-} = I_{c1} - I_{c2}
\end{cases}
\end{equation*}
Note that when the Josephson critical currents I$_{c1}$ and I$_{c2}$ are equal, I$_{c-} = 0$  and the equation \ref{eqn:potential_symm} is retrieved. \newline
The Fig. \ref{fig:fig_asymm} A shows the potential shape for $\phi_+ = \pi /2 $, $\phi_- = 0 $ and several values of I$_{c-}\:\in[-1,1] \; \mu$A. The presence of I$_{c-}\neq 0 $ leads to a shift in the potential minimum and consequently, the wavefunction acquires an offset with respect to $\varphi = 0$ in absence of the external input tone. If a flux-switch is applied, the wavefunction is distributed mostly in the well located on the same wavefunction side with respect to $\varphi = 0$. This makes the fidelity state-dependent and worsens the detector performance.  \newline
The protocol described above has been simulated with QuTip \cite{QuTip} in Fig. \ref{fig:fig_asymm} B (1) and (2). The numerical analysis has been performed by considering I$_{c+}$ = 9 $\mu$A and I$_{c+}$/I$_{c-}$ = $\pm$ 11 $\%$, corresponding to a variation of 20 $\%$ in the critical current value, which is generally larger than the typical fabrication spread. In both simulations, the potential is flipped at t = t$_1$ and the wavefunction collapses always in the same well with a probability of 99.999 $\% $. However, state-independent fidelity can be achieved by adjusting the value of  $\phi_- $. This is shown in Fig. \ref{fig:fig_asymm} parts 3 and 4, where, in the same conditions of simulations reported in Fig. \ref{fig:fig_asymm} B (1) and (2), the state-independent probability has been restored by applying $\phi_m  \simeq \pm 0.46$rad. \newline
The numerical analysis performed with QuTip is supported by classical simulations made in PSCAN2 \cite{pscan2} shown in  Fig. \ref{fig:fig_asymm} C. Sweeping on $\phi_{+}$ and $\phi_{-}$, we have specific regions in which the JDPD is capable of correctly performing phase detection. Simulations have been carried out considering the same parameters as adopted for QuTip, i.e I$_{c+}$ = 9 $\mu$A and I$_{c+}$/I$_{c-}$ = $\pm$ 11 $\%$. In particular, the top part in Fig. \ref{fig:fig_asymm} C refers to the case in which I$_{c+}$/I$_{c-}$ = +11 $\%$, the central part corresponds to  perfect symmetrical JDPD and the bottom part shows the condition when I$_{c+}$/I$_{c-}$ = -11 $\%$. \newline
These considerations demonstrate that the proposed detector is robust with respect to asymmetries that can be corrected by properly changing $\phi_{+}$ and $\phi_{-}$ using dc biases.
\end{section}
\end{document}